# The use of the teleparallelism connection in continuum mechanics

D. H. Delphenich
Kettering, OH USA 45440

**Abstract.** The geometry of parallelizable manifolds – i.e., teleparallelism − is summarized in the language of local frame fields. Some problems in continuum mechanics that relate to the couple-stresses that are produced in the bending and twisting of prismatic beams and wires are then discussed. It is then shown that by going to a higher-dimensional analogue of the geometry that one used for one-dimensional deformable objects, one is basically using the methods of teleparallelism in the context of the Cosserat approach to deformable bodies.

**1. Introduction.** Although differential geometry, as it is defined by the Riemann-Cartan formalism, uses the terms "torsion" and "curvature," nevertheless, when one attempts to formulate continuum mechanics in the language of differential geometry, one eventually notices that the way that continuum mechanics uses the those two geometric terms – for instance, in the study of the bending and twisting of prismatic beams – is not as closely related to the Riemann-Cartan definitions as one would wish. In particular, torsion, in the continuum-mechanical sense, amounts to a one-parameter family of *rotations* about the longitudinal axis of the beam, while torsion, in the Riemann-Cartan sense, relates to infinitesimal *translations* that are associated with the parallel displacement of tangent vectors and frames. Similarly, what constitutes a "curved" object in continuum mechanics does not have to be curved in the Riemann-Cartan sense. For instance, in continuum mechanics one would prefer to think that anything that is not straight or planar is curved, but the Riemann-Cartan curvature of bent wires, cylinders, and cones is zero, since that kind of curvature is more related to the integrability of parallel translation than it is to the more casual, familiar usage of the term. Hence, if one wished to study the work that is required to make the latter deformations and the stresses that are associated with them then the Riemann-Cartan machinery would be of no use.

However, one does find that the continuum-mechanical usage of the words "torsion" and "curvature" is closely related to their earlier usage in the context of moving frames along curves, such as the Frenet-Serret equations. However, in that context, the fundamental object – viz., an anti-symmetric matrix that is associated with an orthonormal frame field along a curve – seems to be more akin to a connection 1-form than it is to a torsion or curvature 2-form.

As we shall show in this article, if one continues along the Frenet-Serret direction to frame fields on higher-dimensional objects than curves then one will find that what one is dealing with is essentially the geometry of "teleparallelism," which is a type of geometry that is defined by a frame field, either locally or globally. A connection is defined by any frame field by postulating that the frame field is parallel, by definition. A vector field or covector field is then parallel with respect to that connection iff its components with respect to the chosen frame field are constant, and, in fact, that property will be true for any other frame field that differs from the original one by a constant transition function; i.e., all of the frames are subjected to the same linear invertible transformation.

If one computes the torsion and curvature 2-forms (in the Riemann-Cartan sense) for the teleparallelism connection then one will find that the torsion 2-form is essentially the exterior derivative of the coframe field that is reciprocal to the frame field, and the curvature is zero. Hence, the vanishing of the torsion 2-form is a necessary (and locally sufficient) condition for the integrability of the coframe field into a "natural" coframe field; i.e., one of the form $dx^i$, $i = 1, \ldots, n$ for some coordinate chart $(U, x^i)$ that is contained in the support of the frame field. The vanishing of the curvature means that the parallel translation of a tangent vector or linear frame around a loop will always bring it back to itself. However, if one uses *affine* tangent vectors and frames then the torsion will contribute a translation of the point of application of the vector or frame.

Conventional continuum mechanics does not generally consider the association of tangent frames to deformable bodies, but only diffeomorphisms of the region of space that the body occupies. Thus, the conventional – i.e., Cauchy-Green – way of defining strain considers only what one might call *metric strain*. That is, one considers the changes in the distances between points of an object under deformation.

However, there is a broader picture for the deformation of material objects that goes back to the work of Eugene and François Cosserat, who had hoped to apply Darboux's theory of moving frames to the study of deformation in continuum mechanics. In fact, specialized forms of their general theory had existed prior to the publication of their *magnum opus* [1] in the work done by Waldemar Voigt on the elasticity of crystals [2] and Lord Kelvin and Peter Tait's monumental treatise on natural philosophy [3], among others.

Basically, what distinguishes a Cosserat medium from a conventional one is that every point of the medium is associated with a space of oriented, orthonormal frames; i.e., it is the bundle of oriented, orthonormal frames on the object. Hence, in addition to the metric strain that pertains to the deformation of the points, one can also consider *frame strain*, which pertains to the deformation of the frames at the various points. This becomes physically meaningful when one can define a notion of work that is associated with the deformation of frames, which also implies the existence of "couple-stresses" in the medium. For instance, one knows that it will take work to bend or twist a flexible wire, even when it is inextensible. In such a case, the metric strain that is associated with the deformation of the points of the wire is less useful than the frame strain that is associated with the deformation of a frame field along the wire.

In this article, we shall first briefly summarize some of the key aspects of the geometry of parallelizable manifolds, which are actually somewhat rare, although we shall define things only for local frame fields, which always exist. We will then discuss how the geometry of moving frames along curves relates to the problems of the bending and twisting of elastic wires, and introduce the Cosserat equations for equilibrium in that context. We will then show that when one goes to higher-dimensional deformable objects than one-dimensional ones, one will find that one is basically looking at the geometry of teleparallelism, and one can regard the basic congruences of an embedding as moving frames along curves whose equations are the equations of teleparallelism, when restricted to the individual parameters.

**2. Teleparallelism** ([1]). If one takes the position that what a connection on the tangent bundle $T(M) \to M$ or bundle $GL(M) \to M$ of linear frames on an $n$-dimensional differentiable manifold $M$ is supposed to accomplish is to allow one to compare tangent vectors or frames at finitely-separated points then there will be two basic ways to do that:

The most widely accepted approach, namely, the Riemann-Cartan approach, is to introduce an infinitesimal linear connection, which can take the form of a 1-form $\omega_j^i$ on $GL(M)$ that takes its values in the Lie algebra $\mathfrak{gl}(n)$, which consists of infinitesimal generators of one-parameter families of linear transformations of tangent frames, and then define the parallel translation of frames along curves. In particular, if $\mathbf{e}_i(s)$ is a frame field along a curve $x(s)$ in $M$ whose velocity vector field is $\mathbf{v}(s)$ then it will be *parallel* for the connection 1-form $\omega_j^i$ iff:

$$\frac{d\mathbf{e}_i}{ds} = \mathbf{e}_j \omega_i^j (\mathbf{v}(s)). \tag{2.1}$$

Thus, as long as two distinct points $x_0$ and $x_1$ lie in a path-connected component of $M$ one can compare their tangent spaces and say that a frame at $x_0$ is parallel to a frame at $x_1$ relative to a chosen curve iff its parallel translate along that curve agrees with the frame at $x_1$. However, the parallel translate of the initial frame can depend considerably upon the choice of curve, and, in fact, it is the non-vanishing of the curvature of a linear connection that obstructs the path-independence of parallel translation. As a consequence, if the curvature of $\omega_j^i$ is non-vanishing at a point of $M$ then there will not even be a local neighborhood of that point that has a local frame field defined on it whose members are all parallel vector fields for $\omega_j^i$.

The other elementary possibility for comparing tangent objects is to define a local frame field $\mathbf{e}_i(x)$ on some neighborhood $U$ of the given point and *define* it to be a parallel frame field. Thus, two tangent objects (i.e., tensors) at distinct points of $U$ will be parallel [with respect to $\mathbf{e}_i(x)$] iff they have the same components with respect to the frames of $\mathbf{e}_i(x)$ at those respective points.

Parallel translation of tangent objects to the other points of $U$ is then path-independent and one says that a vector field $\mathbf{v}(x) = v^i(x) \mathbf{e}_i(x)$ on $U$ is *parallel* iff its components $v^i(x)$ are constants; i.e., iff:

$$dv^i = 0. \tag{2.2}$$

Since frame fields are composed of vector fields, one then sees that one can also define a frame field $\mathbf{f}_i(x)$ on $U$ to be *parallel* with respect to $\mathbf{e}_i(x)$ iff the transition function $L_i^j(x)$ on $U$ that takes $\mathbf{e}_i(x)$ to $\mathbf{f}_i(x)$ by way of:

$$\mathbf{f}_i(x) = \mathbf{e}_j(x) L_i^j(x) \tag{2.3}$$

is also constant; i.e., iff:

---

[1] Some good textbooks on differential geometry for the geometry of parallelizable manifolds are Bishop and Crittenden [**4**] and Sternberg [**5**]. In particular, the former reference introduces it in the form of a series of problem sets.

$$dL_i^j = 0. \tag{2.4}$$

This also means that every vector field that is parallel with respect to $\mathbf{e}_i(x)$ will be parallel with respect to $\mathbf{f}_i(x)$, as well. Thus, one can define an equivalence class of frame fields on $U$ that produce the same definition of parallelism for vector fields.

One can also define the parallelism of covector fields by introducing the coframe field $\theta^i|_x$ on $U$ that is reciprocal to $\mathbf{e}_i(x)$, namely, the one that makes:

$$\theta^i|_x(\mathbf{e}_j(x)) = \delta_j^i \tag{2.5}$$

for every $x \in U$.

One then says that $\theta^i$ is also parallel, so a covector field $\alpha = \alpha_i(x)\, \theta^i$ will be *parallel* on $U$ iff its components $\alpha_i(x)$ are constants; i.e., iff:

$$d\alpha_i = 0. \tag{2.6}$$

One obtains an equivalence class of coframe fields on $U$ that are all parallel to each other in a manner that is similar to the one above for frame fields.

One can associate a 1-form $\omega_j^i$ on $U$ with values in $\mathfrak{gl}(n)$ that will represent the connection 1-form for the definition of parallelism by using the equation of parallelism for the frame field $\mathbf{e}_i(x)$:

$$d\mathbf{e}_i = \mathbf{e}_j \otimes \omega_i^j. \tag{2.7}$$

If $U$ admits a coordinate system $x^i$, $i = 1, \ldots, n$ then one can define a *natural* frame field on $U$ by way of the vector fields $\partial_i = \partial / \partial x^i$ and a natural coframe field by means of the covector fields $dx^i$. If one expresses $\mathbf{e}_i(x)$ in the form ([1]):

$$\mathbf{e}_i = \partial_j\, \tilde{h}_i^j \tag{2.8}$$

then one will have:

$$d\mathbf{e}_i = \partial_j \otimes d\tilde{h}_i^j = \mathbf{e}_j \otimes h_k^j\, d\tilde{h}_i^k, \tag{2.9}$$

so one can set:

$$\omega_i^j = h_k^j\, d\tilde{h}_i^k \tag{2.10}$$

for the local representative of $\omega$ with respect to the natural frame field.

However, connection 1-forms are not tensor fields, so when one wishes to find the local representative with respect to the given frame field $\mathbf{e}_i(x)$, one must use the transformation formula ([2]):

$$\omega' = \tilde{h}\omega h + \tilde{h}dh = d\tilde{h}\,h + \tilde{h}dh = d(\tilde{h}h) = 0, \tag{2.11}$$

since $\tilde{h}$ is the inverse of $h$.

---

([1]) Here, and in what follows, a tilde over a matrix will denote its inverse.
([2]) When no confusion will arise, we shall suppress the indices on matrices.

This is consistent with the fact that a vector field **v** will be parallel iff:

$$0 = \nabla \mathbf{v} \equiv \nabla v^i \otimes \mathbf{e}_i , \qquad (2.12)$$

in which we have introduced the *covariant differential* $\nabla \mathbf{v}$ of the vector field **v** and set:

$$\nabla v^i = dv^i + \omega^i_j v^j . \qquad (2.13)$$

However, we also wish that **v** should be parallel iff $dv^i = 0$, which demands that we must have $\omega^i_j v^j = 0$, as well. Since this must be true for all parallel vector fields, that would imply that $\omega^i_j = 0$, relative to $\mathbf{e}_i$; i.e.:

$$\nabla \mathbf{v} = dv^i \otimes \mathbf{e}_i ; \qquad (2.14)$$

this is, of course, consistent with (2.11).

If one wishes to consider $\nabla \mathbf{v}$ relative to the natural frame field then one must use $\omega^i_j$ as it is defined in (2.10). If $\mathbf{v} = \bar{v}^i \partial_i$ then one will get:

$$\nabla \mathbf{v} = (d\bar{v}^i + \omega^i_j \bar{v}^j) \otimes \partial_i . \qquad (2.15)$$

One has similar constructions and results for a covector field $\alpha = \alpha_i \theta^i = \bar{\alpha}_i dx^i$:

$$\nabla \alpha = d\alpha_i \otimes \theta^i = (d\bar{\alpha}_i - \omega^j_i \bar{\alpha}_j) \otimes dx^i. \qquad (2.16)$$

It is straightforward to define the geodesics in $U$ relative to this connection $\omega^i_j$, which we shall refer to as the *teleparallelism connection*, since it played such a fundamental role in one of Einstein's early attempts to unify the theories of gravitation and electromagnetism ([1]). Namely, a curve $x(s)$ in $U$ is a *geodesic* iff its velocity vector field is parallel along $x(s)$ with respect to the chosen frame field $\mathbf{e}_i(x)$. Hence, its components with respect to $\mathbf{e}_i(x)$ will be constants, while its components with respect to $\partial_i$ will satisfy:

$$0 = \nabla_\mathbf{v} \bar{v}^i = \frac{d\bar{v}^i}{ds} + \omega^i_j(\mathbf{v}) \bar{v}^j = \frac{d\bar{v}^i}{ds} + \omega^i_{jk} \bar{v}^j \bar{v}^k , \qquad (2.17)$$

in which $\omega^i_j = \omega^i_{jk} dx^k$.

If $\mathbf{e}_i(x)$ were $\partial_i$ then the constancy of the velocity components would make a geodesic a straight line in that coordinate system. Hence, one can say that in the general case, a geodesic is a curve that looks like a straight line to the chosen frame field.

---

([1]) The author has compiled an anthology [**6**] of English translations of many of the early papers on teleparallelism that is available as a free download at his website: www.neo-classical-physics.info.

One can obtain the torsion and curvature 2-forms for $\omega^i_j$ most directly by using the form that it takes with respect to $\mathbf{e}_i(x)$, namely, $\omega^i_j = 0$. The form that the Cartan structure equations take on $GL(U)$ is:

$$\Theta^i = d_\wedge \theta^i + \omega^i_j \wedge \theta^j, \qquad \Omega^i_j = d_\wedge \omega^i_j + \omega^i_k \wedge \omega^k_j, \qquad (2.18)$$

in which we are using the symbol $d_\wedge$ for the exterior derivative operator, in order to not confuse it with the differential of the same tensor fields.

When one pulls the equations (2.18) down to $U$ by way of the frame field $\mathbf{e}_i(x)$, they become simply:

$$\Theta^i = d_\wedge \theta^i, \qquad \Omega^i_j = 0. \qquad (2.19)$$

The fact that the curvature vanishes should have been expected from the existence of a parallel frame field on $U$. However, the torsion 2-form does not have to vanish, and one sees that it will vanish iff the coframe field $\theta^i$ (and therefore the frame field $\mathbf{e}_i$) is, by definition, *holonomic*; in the contrary case, it is said to be *anholonomic*. Any natural frame field is clearly holonomic, and, in fact, by using the Poincaré lemma and the inverse function theorem, one can show that the converse is locally true.

If one puts the first structure equation into the form that it takes in the given coframe field $\theta^i$, namely:

$$\Theta^i = \tfrac{1}{2} S^i_{jk} \theta^j \wedge \theta^k, \qquad (2.20)$$

then one can interpret the components in terms of the *structure functions* $c^i_{jk}(x)$ of $\theta^i$ ( or $\mathbf{e}_i$). They are defined by either:

$$[\mathbf{e}_j, \mathbf{e}_k] = c^i_{jk}(x)\, \mathbf{e}_i \qquad \text{or} \qquad d_\wedge \theta^i = -\tfrac{1}{2} c^i_{jk}(x)\, \theta^j \wedge \theta^k. \qquad (2.21)$$

One gets from the first form to the second one most directly by using the "intrinsic" formula for the exterior derivative of a 1-form $\alpha$:

$$d_\wedge \alpha(\mathbf{v}, \mathbf{w}) = \mathbf{v}(\alpha(\mathbf{w})) - \mathbf{w}(\alpha(\mathbf{v})) - \alpha[\mathbf{v}, \mathbf{w}]. \qquad (2.22)$$

Thus, one sees that the components of the torsion 2-form relative to the parallel frame field amount to:

$$S^i_{jk} = -\, c^i_{jk}. \qquad (2.23)$$

Hence, the torsion of the teleparallelism connection vanishes iff the frame field that defines it is holonomic. This partially explains why some early researchers in teleparallelism were characterizing it as "anholonomic Euclidian geometry," although we shall have to also show that the connection is a metric connection, which we shall do shortly.

If one puts the first structure equation into the form that it takes in the natural coframe field, namely:

$$\Theta^i = \omega^i_j \wedge dx^j = -\tfrac{1}{2}(\omega^i_{jk} - \omega^i_{kj})\, dx^j \wedge dx^k, \qquad (2.24)$$

then one will see that the components of torsion relative to that frame field are minus the anti-symmetric part of the connection 1-form, which is how one often finds it defined in mainstream relativity theory.

When one has a local frame field $\mathbf{e}_i(x)$ on $U$, with a reciprocal coframe field $\theta^i$, one can define some other geometrically fundamental tensor fields, in addition to the connection 1-form. First of all, one can define a volume element $V$ (i.e., a non-zero $n$-form in $U$) by demanding that volume of the parallelepiped that is defined by any frame $\mathbf{e}_i(x)$ should be unity. This makes:

$$V = \theta^1 \wedge \ldots \wedge \theta^n = \frac{1}{n!} \varepsilon_{i_1 \cdots i_n} \theta^{i_1} \wedge \cdots \wedge \theta^{i_n}. \qquad (2.25)$$

One immediately sees that since its components are constants it is parallel for the teleparallelism connection; i.e.:

$$\nabla V = 0. \qquad (2.26)$$

Thus, the connection 1-form $\omega^i_j$ takes its values in $\mathfrak{sl}(n)$, and parallel translation preserves the orientation of frames.

Furthermore, one can define a metric $g$ (of any signature type) on $U$ by demanding that $\mathbf{e}_i(x)$ must be orthonormal for that metric. If we use the Euclidian signature type then that will make $g(\mathbf{e}_i, \mathbf{e}_j) = \delta_{ij}$, by definition, and the metric tensor $g$ will then take the form:

$$g = \delta_{ij}\, \theta^i\, \theta^j, \qquad (2.27)$$

in which the product on the right-hand side is the symmetrized tensor product of the covector fields.

Once again, since the components of the metric tensor field are constants with respect to $\theta^i$, the metric will be parallel for the teleparallelism connection; i.e.:

$$\nabla g = 0. \qquad (2.28)$$

Hence, it will be a metric connection for that metric, so $\omega^i_j$ will take its values in $\mathfrak{so}(n)$, and parallel translation will preserve the scalar product of tangent vectors.

Thus, we see that, in fact, the geometry that one sees with respect to the frame field $\mathbf{e}_i$ will be essentially Euclidian, although it is only when the frame field is holonomic that one is truly dealing with Euclidian geometry.

So far, we have only defined frame fields on open subsets of the manifold $M$, whereas classically the research on teleparallelism defined them on all of $M$. Actually, it is a very deep topological problem whether a differentiable manifold truly admits a global frame field. In such a case, one calls it *parallelizable*. Although every Lie group is parallelizable, even such homogeneous spaces as most spheres (except for the ones in dimension 0, 1, 3, 7) are not parallelizable. Indeed, the first definitive work on the subject of the topological obstructions to parallelizability did not appear until some years after the physicists had given up on teleparallelism as a path to unification, namely, the doctoral thesis [**7**] of Edward Stiefel, who defined what are now called *Stiefel-Whitney classes* for a manifold, and take the form of cocycles with values in $\mathbb{Z}_2$. Of course, if one also considers that all of the early researchers in teleparallelism regarded manifolds by starting with coordinate systems, which are generally defined only locally, in effect, they were not really defining frame fields on all of spacetime, anyway.

**3. The deformable line.** We shall confine ourselves to the non-relativistic arena, so the space in which material objects and deformations will take place will be three-dimensional Euclidian space $E^3 = (\mathbb{R}^3, \delta)$, where $\delta$ is our notation for the Euclidian scalar product; e.g., $\delta(\mathbf{v}, \mathbf{w})$.

*a. Oriented, orthonormal frames on $E^3$.* The canonical basis for $\mathbb{R}^3$ will be denoted by the column vectors:

$$\boldsymbol{\delta}_1 = (1, 0, 0)^T, \qquad \boldsymbol{\delta}_2 = (0, 1, 0)^T, \qquad \boldsymbol{\delta}_3 = (0, 0, 1)^T, \qquad (3.1)$$

and its reciprocal basis $\{\delta^i, i = 1, 2, 3\}$ for the dual space $\mathbb{R}^{3*}$ will be defined by the corresponding row vectors.

This basis (i.e., 3-frame) is then orthonormal for the Euclidian scalar product:

$$\delta(\boldsymbol{\delta}_i, \boldsymbol{\delta}_j) = \delta_{ij}, \qquad (3.2)$$

as is the reciprocal basis.

Generally, we will use the corresponding frame field $\{\partial_i, i = 1, 2, 3\}$ and reciprocal coframe field $\{dx^i = 1, 2, 3\}$ that are defined by using the components $x^i$ of a vector $\mathbf{x} = x^i \boldsymbol{\delta}_i$ as the coordinates of the point that is described by its head, and defining the frame field that is obtained from the directional derivative operators along the coordinate directions:

$$\partial_i \equiv \frac{\partial}{\partial x^i}. \qquad (3.3)$$

Thus, a typical vector field $\mathbf{v}(x)$ and covector field $\alpha_x$ on $E^3$ will take the forms:

$$\mathbf{v}(x) = v^i(x)\, \partial_i, \qquad \alpha_x = \alpha_i(x)\, dx^i, \qquad (3.4)$$

in which the component functions $v^i(x)$, $\alpha_i(x)$ will be assumed to be smooth, or at least continuously-differentiable as many times as required by the application.

The bundle $SO(E^3) \to E^3$ of oriented, orthonormal frames on $E^3$ is trivial, so it can be represented in the form $E^3 \times SO(3)$, and a typical element $(x, \mathbf{e}_i)$ − so $\mathbf{e}_i$ is an oriented, orthonormal frame at the point $x \in E^3$ – can be represented by coordinates $(x^i, R_i^j)$, where:

$$\mathbf{e}_i = \partial_j R_i^j, \qquad (3.5)$$

and the matrix $[R] = R_i^j$ is orthogonal with a unity determinant:

$$[R]^T[R] = [R][R]^T = I, \ \det[R] = 1. \qquad (3.6)$$

Thus, it represents a three-dimensional Euclidian rotation that preserves the orientation of any frame; i.e., a proper rotation.

The bundle $SO^*(E^3)$ of oriented, orthonormal coframes is likewise trivial, and the typical element $(x, \theta^i)$ can be represented by coordinates $(x^i, R_j^i)$, where this time:

$$\theta^i = R_j^i \, dx^j. \qquad (3.7)$$

Sections of the two bundles are then global fields of oriented, orthonormal frames and coframes, respectively, which then take the forms:

$$\mathbf{e}_i(x) = \partial_j \tilde{R}_i^j(x), \qquad \theta_x^i = R_j^i(x) \, dx^j, \qquad (3.8)$$

respectively. (The reason that we are using the inverse of $R$ in the first of these systems of equations is that one must ensure that when $\theta_x^i$ is the reciprocal coframe field to $\mathbf{e}_i(x)$, one will actually have $\theta_x^i(\mathbf{e}_j) = \delta_j^i$.)

*b. Oriented, orthonormal frame fields along curves in $E^3$.* A smooth curve $x: \mathbb{R} \to E^3$, $s \mapsto x(s)$ will be defined by its coordinate functions $x^i(s)$, relative to the canonical coordinate system. Because we are only dealing with the affine space $E^3$, we can identify all successive derivatives of $x(s)$ with respect to $s$ with vector fields on $x(s)$:

$$\mathbf{v}(s) = \frac{dx}{ds} = \left.\frac{dx^i}{ds}\right|_s \partial_i, \qquad \mathbf{a}(s) = \frac{d^2x}{ds^2} = \frac{d\mathbf{v}}{ds} = \left.\frac{d^2x^i}{ds^2}\right|_s \partial_i = \left.\frac{dv^i}{ds}\right|_s \partial_i, \qquad (3.9)$$

and one typically refers to these vector fields as the *velocity* and *acceleration* of the curve, even when $s$ does not play the role of time, such as when it represents arc length.

The parameterization of the curve will be said to have *unit speed* when:

$$\|\mathbf{v}(s)\|^2 = \delta(\mathbf{v}(s), \mathbf{v}(s)) = 1. \qquad (3.10)$$

As a consequence, the acceleration of such a curve will always satisfy:

$$\delta(\mathbf{v}(s), \mathbf{a}(s)) = 0. \qquad (3.11)$$

Thus, it will either be zero or always orthogonal to the velocity.

It if it is zero then the curve $x(s)$ will be a straight line – say, $a^i + sb^i$, where $a^i$ and $b^i$ are constants. Thus, the acceleration is also a measure of the "curvature" of the curve, in the older (i.e., Frenet-Serret [**8**]) sense of the word.

If $\mathbf{a}(s) \neq 0$ then one can express it as the product:

$$\mathbf{a}(s) = \kappa(s)\, \mathbf{e}_2(s) \tag{3.12}$$

of a scalar function $\kappa(s) = \|\mathbf{a}(s)\|$ that one calls the *curvature* of the curve $x(s)$ and a unit vector field $\mathbf{e}_2(s)$ along $x(s)$ that gives its direction, and which one calls the *principal normal* vector field for the curve:

If we assume that $x(s)$ has a unit-speed parameterization then $\mathbf{v}(s)$ will be a unit vector field along the curve and we can rename it $\mathbf{e}_1(s)$, which is then the *unit tangent* vector field along the curve. From (3.9) and (3.12), we then have:

$$\frac{d\mathbf{e}_1}{ds} = \kappa(s)\, \mathbf{e}_2(s). \tag{3.13}$$

(When the curve is a straight line, one will have $\kappa(s) = 0$.)

Since $\mathbf{e}_2(s)$ will be orthogonal to $\mathbf{e}_1(s)$ for all $s$, and both are unit vector fields, we can complete the right-handed, orthonormal triad with the *binormal* vector field:

$$\mathbf{e}_3(s) \equiv \mathbf{e}_1(s) \times \mathbf{e}_2(s). \tag{3.14}$$

This particular frame field along $x(s)$ is adapted to the curve [i.e., $\mathbf{e}_1(s) = \mathbf{v}(s)$], and is called the *Frenet frame field* along the curve, when the curve admits one. The only other case is the straight line, so curves that admit Frenet frame fields are, in a sense, "generic."

One then finds that eq. (3.13) can be augmented with two others that give the equations of the moving frame along $x(s)$:

$$\frac{d\mathbf{e}_i}{ds} = \mathbf{e}_j(s)\, \omega_i^j(s), \tag{3.15}$$

in which:

$$\omega_i^j(s) = \begin{bmatrix} 0 & \kappa(s) & 0 \\ -\kappa(s) & 0 & \tau(s) \\ 0 & -\tau(s) & 0 \end{bmatrix} = \tau I_1 + \kappa I_3, \tag{3.16}$$

which are then the *Frenet-Serret* equations for the Frenet frame field. In (3.16), we have introduced the elementary basis matrices for $\mathfrak{so}(3)$:

$$I_1 = \begin{bmatrix} 0 & 0 & 0 \\ 0 & 0 & 1 \\ 0 & -1 & 0 \end{bmatrix}, \qquad I_2 = \begin{bmatrix} 0 & 0 & -1 \\ 0 & 0 & 0 \\ 1 & 0 & 0 \end{bmatrix}, \qquad I_3 = \begin{bmatrix} 0 & 1 & 0 \\ -1 & 0 & 0 \\ 0 & 0 & 0 \end{bmatrix}, \tag{3.17}$$

which describe infinitesimal rotations about the *x*, *y*, and *z* axes, respectively; i.e., **e**$_1$(*s*), **e**$_2$(*s*). **e**$_3$(*s*), resp.

The other parameter $\tau(s)$ that is introduced is referred to as the *torsion* of the curve. It vanishes iff the curve lies in the plane of **e**$_1$(*s*) and **e**$_2$(*s*) (viz., the *osculating plane* to the curve), which will then be the same plane for all *s*. Indeed, one sees that the elementary matrix $I_1$ describes an infinitesimal rotation in the plane that is orthogonal to **e**$_1$(*s*); i.e., a twisting motion about the velocity vector field whose rate of rotation is the torsion.

Similarly, the elementary matrix $I_3$ describes an infinitesimal rotation of the frame field about the binormal whose rate of rotation is the curvature.

More generally, if one has an adapted, oriented, orthonormal frame field **e**$_i$(*s*) along *x*(*s*) then the equations of that frame field will still take the form of the system in (3.16), except that $\omega_i^j(s)$ will take the form:

$$\omega_i^j(s) = \begin{bmatrix} 0 & \kappa(s) & -\lambda(s) \\ -\kappa(s) & 0 & \tau(s) \\ \lambda(s) & -\tau(s) & 0 \end{bmatrix} = \tau I_1 + \lambda I_2 + \kappa I_3 , \qquad (3.18)$$

in which $\lambda(s)$ amounts to a rate of rotation about the principal normal **e**$_2$(*s*); i.e., in the plane of the tangent and binormal. In such a case, **e**$_2$(*s*) will no longer have to represent the acceleration of the curve, although the oriented, orthonormal frame field{**e**$_2$(*s*), **e**$_2$(*s*)} will still span the plane that is normal to the velocity vector field **e**$_1$(*s*).

If one represents the general oriented, orthonormal frame field **e**$_i$(*s*) along *x*(*s*) in the form (3.8) then one will have:

$$\frac{d\mathbf{e}_i}{ds} = \partial_j \frac{d\tilde{R}_i^j}{ds} = \mathbf{e}_j R_k^j \frac{d\tilde{R}_i^k}{ds}, \qquad (3.19)$$

which makes:

$$\omega_i^j(s) = R_k^j \frac{d\tilde{R}_i^k}{ds}. \qquad (3.20)$$

If *s* plays the role of time then this matrix will represent the angular velocity of the frame field **e**$_i$(*s*) along *x*(*s*), relative to itself, while $d\tilde{R}_i^j / ds$ will represent the angular velocity relative to the holonomic frame field $\partial_i$.

The relationship between the equations of motion or deformation of a frame field along a curve and the teleparallelism connection are rather suggestive at this point, since equation (3.20) looks like a one-parameter version of the equation (2.10) for the teleparallelism connection 1-form. In the next section, we shall see that when one has a frame field on a congruence of curves, one can regard equations of the form (3.20) as simply the pull-backs of the equation that defines the teleparallelism 1-form along the embedding that gives the congruence.

*c. Statics of the deformable line.* In Kelvin and Tait ([**3**], vol. II), one finds that their treatment of the deformable line amounts to first regarding its kinematical state of deformation as being defined by the independent components of $\omega_i^j(s)$, which were then arranged into a vector ($\tau(s)$, $\lambda(s)$, $\kappa(s)$). Thus, they were actually considering the deformation of the frame field on $x(s)$, more than the deformation of the curve itself.

In fact, to be precise, the vector ($\tau(s)$, $\lambda(s)$, $\kappa(s)$) [or, equivalently, the matrix $\omega_i^j(s)$] is not associated with a *deformation* of the curve, but with a *state* of the curve, unless one chooses the initial state of the curve before deformation to be a straight line, for which one would have $\omega_i^j(s) = 0$. Thus, in order to characterize the deformation of one curve to another − or rather, one frame field along a curve to another – one really needs to consider how $\omega_i^j(s)$ itself will change from the initial state to the final one.

Thus, imagine that a curve exists in an initial state $x(s)$ and a deformed state $\bar{x}(s)$, along with an initial oriented, orthonormal frame field $\mathbf{e}_i(s)$ on $x(s)$ and a deformed one $\bar{\mathbf{e}}_i(s)$ on $\bar{x}(s)$. The equations of $\mathbf{e}_i(s)$ take the form (3.16) with $\omega_i^j(s)$ taking either the general form (3.18) or the Frenet-Serret form (3.16), while the equations of $\bar{\mathbf{e}}_i(s)$ are:

$$\frac{d\bar{\mathbf{e}}_i}{ds} = \bar{\mathbf{e}}_j \, \bar{\omega}_i^j. \tag{3.21}$$

If the deformation of the frame field $\mathbf{e}_i(s)$ into $\bar{\mathbf{e}}_i(s)$ takes the form:

$$\bar{\mathbf{e}}_i(s) = \mathbf{e}_j(s) \tilde{R}_i^j(s) \tag{3.22}$$

then the differentiation in (3.21) will give:

$$\dot{\mathbf{e}}_j \tilde{R}_i^j + \mathbf{e}_j \dot{\tilde{R}}_i^j = \mathbf{e}_j \omega_k^j \tilde{R}_i^k + \mathbf{e}_j \dot{\tilde{R}}_i^j = \bar{\mathbf{e}}_j (\eta_k^j + R_k^j \omega_l^k \tilde{R}_i^l) = \bar{\mathbf{e}}_j \bar{\omega}_i^j, \tag{3.23}$$

in which we have introduced:

$$\eta_i^j = R_k^j \dot{\tilde{R}}_i^k = \bar{\omega}_i^j - R_k^j \omega_l^k \tilde{R}_i^l. \tag{3.24}$$

This last equation can be regarded as either the formula for the transformation of a connection from one frame field to another or the definition of a measure of the change in the connection under the deformation, as viewed from the deformed frame. If one uses the Frenet interpretation of the elements of the two connection matrices then one can also say that it represents the change in the torsion and curvature of the line as a result of the deformation. Thus, we are justified in using:

$$\eta_i^j = \begin{bmatrix} 0 & \Delta\kappa & -\Delta\lambda \\ -\Delta\kappa & 0 & \Delta\tau \\ \Delta\lambda & -\Delta\tau & 0 \end{bmatrix} \tag{3.25}$$

as a measure of the finite strain that is associated with the deformation of the frame fields. We shall then call the 1-form $\eta_i^j \, ds$ with values in $\mathfrak{so}(3)$ the *finite frame strain 1-form* for the deformation.

Of course, when one starts with a straight line, $\omega$ will be zero, and $\eta$ will coincide with the deformed connection matrix $\bar{\omega}$, which is essentially the situation that was described in Kelvin and Tait. However, we shall wish to associate work with the *deformation* of a frame field, not the frame field itself.

In the calculus of variations, one also needs to know what would constitute an infinitesimal deformation of $\omega$, and the way that one arrives at such a notion is by starting with a *finite variation* of the frame field $\mathbf{e}_i(s)$ along $x(s)$. This will take the form of a differentiable two-parameter family $R(s, h)$ of rotations such that $h$ ranges from 0 to some unspecified finite value and $R(s, 0)$ gives the identity. A finite variation of the frame field $\mathbf{e}_i(s)$ will then take the form:

$$\Delta \mathbf{e}_i(s, h) = \mathbf{e}_i(s)\, \tilde{R}(s, h). \tag{3.26}$$

In order to pass to the infinitesimal generator of such a finite variation, one differentiates with respect to $h$ and evaluates at $h = 0$. The result is a *variation* – or *virtual displacement* – of the frame field along the curve:

$$\delta \mathbf{e}_i(s) \equiv \left.\frac{\partial(\Delta \mathbf{e}_i)}{\partial h}\right|_{h=0} = \mathbf{e}_i(s)\, \delta \tilde{R}(s), \tag{3.27}$$

in which:

$$\delta \tilde{R}(s) \equiv \left.\frac{\partial \tilde{R}(s,h)}{\partial h}\right|_{h=0} \tag{3.28}$$

becomes a smooth function from $\mathbb{R}$ to $\mathfrak{so}(3)$ that we regard as a variation or virtual displacement of $\tilde{R}$.

A virtual displacement of $\mathbf{e}_i(s)$ will be associated with a corresponding variation of $\omega$ by way of the equations of the moving frame. One basically takes the variation of both sides:

$$\delta \frac{d\mathbf{e}_i}{ds} = \frac{d}{ds}(\delta \mathbf{e}_i) = \frac{d\mathbf{e}_i}{ds}\delta\tilde{R} + \mathbf{e}_i \frac{d(\delta\tilde{R})}{ds} = \mathbf{e}_i\left(\omega\,\delta\tilde{R} + \frac{d(\delta\tilde{R})}{ds}\right),$$

$$\delta(\mathbf{e}_i\, \omega) = \delta\mathbf{e}_i\, \omega + \mathbf{e}_i\, \delta\omega = \mathbf{e}_i\, (\delta\tilde{R}\, \omega + \delta\omega),$$

which gives:

$$\delta\tilde{R}\, \omega + \delta\omega = \omega\, \delta\tilde{R} + \frac{d(\delta\tilde{R})}{ds},$$

and solving for $\delta\omega$ gives:

$$\delta\omega = \frac{d(\delta\tilde{R})}{ds} + [\omega, \delta\tilde{R}]. \tag{3.29}$$

This expression for $\delta\omega$ takes the form of a "covariant derivative" of the virtual displacement $\delta\tilde{R}$ when one uses $\omega$ as the connection, and we shall regard $\delta\omega$ as a measure of infinitesimal strain on the torsion and curvature of the moving frame ($x(s)$, $\mathbf{e}_i(s)$).

The stress that is produced by the deformation of a frame field on a material line takes the form of a *couple-stress*, as one finds in the Cosserat approach to deformation. The couple-stress matrix $\mu^i_j(s)$ will be a dual object to the virtual displacement $\delta\omega^j_i(s)$, so since the latter object takes its values in $\mathfrak{so}(3)$, $\mu^i_j(s)$ will take its values in $\mathfrak{so}(3)^*$. The canonical bilinear pairing $\mathfrak{so}(3)^* \times \mathfrak{so}(3)$, $(\mu, \delta\omega) \mapsto \mu(\delta\omega)$ is defined by the *virtual work*:

$$\delta W = \mu^i_j \, \delta\omega^j_i \tag{3.30}$$

that is associated with the virtual displacement $\delta\omega^j_i(s)$ of the initial line.

If the stress admits a stress potential $W(\omega^j_i, \dot\omega^j_i, \ldots)$ then one can also say that:

$$\mu^i_j = \frac{\delta W}{\delta \omega^j_i}, \tag{3.31}$$

which makes $\mu^i_j$ a function of $\omega^j_i$, and possibly its higher derivatives. This functional relationship then takes the form of a mechanical constitutive law for the association of infinitesimal frame strains with corresponding couple stresses. The one that was chosen by Kelvin and Tait was the linear case:

$$\mu^i_j = C^{ik}_{jl} \omega^l_k, \tag{3.32}$$

which they defined more simply by using essentially $\mu_i = (K, L, M)$ for the couple stresses and $\omega^j = (\tau, \lambda, \kappa)$, which relate to the corresponding matrices $\mu^i_j$ and $\omega^j_i$ by the adjoint map, which also takes the form:

$$[\mu] = \mu_i \, I^i, \qquad [\omega] = \omega^j I_j \qquad (I^i = \delta^{ij} I_j = I_i). \tag{3.33}$$

The linear constitutive law for couple stress then takes the form:

$$\mu_i = C_{ij} \, \omega^j. \tag{3.34}$$

One can obtain the equations of equilibrium for the deformable line from using d'Alembert's principle of vanishing virtual work for any allowable virtual displacement or Hamilton's principle of least action. For the case of the flexible, inextensible line, one only considers the couple stresses that are associated with the bending and twisting of the curve, and the equations take the form:

$$0 = \frac{d\mu^i_j}{ds}, \tag{3.35}$$

so when the moments are applied to only the extremities of the line, the effect is to distribute the couple-stresses uniformly along it.

When one also considers the possible elongation or contraction of the line in response to applied forces, there will also be an associated force-stress tensor $\sigma^i_j$, which does not have to be symmetric, in the general case. If one imposes the constraint on the action functional that it be invariant under Euclidian rigid motions, as the Cosserats did, then one will arrives at a different set of equilibrium equations:

$$0 = \frac{d\sigma^0_j}{ds}, \qquad 0 = \frac{d\mu_{ij}}{ds} + \sigma_{0i}\dot{x}_j - \sigma_{j0}\dot{x}_i. \qquad (3.36)$$

The appearance of the extra term in the moment-couple-stress equation is purely an artifact of the invariance of the action, and represents essentially an internal couple-stress that comes about in the absence of applied force and moment densities solely due to the algebraic properties of the group of rigid motions. (For a more detailed discussion of this, one might confer the author's paper [**9**], in addition to the Cosserat references, as well as a paper by Sudria [**10**], in which he revisited the Cosserat argument.)

**4. Generalization to higher dimensions.** Although Kelvin and Tait treated the theory of infinitesimal deformations of planar plates with a different approach from what they had used for deformable lines, it is actually possible to continue to use the Frenet-Serret conception of torsion and curvature for objects embedded in $E^3$ that have more dimensions that just the one.

One finds that one is then regarding the concept of a frame field from the same standpoint as Levi-Civita in the chapter on congruences in his book [**11**] on the "absolute differential calculus."

*a. Frame fields and basic congruences.* Let $x: \mathcal{O} \subset \mathbb{R}^p \to E^3$, $u^a \mapsto x^i(u^a)$ be an embedding of a *p*-dimensional object in $E^3$, where *p* might be 2 or 3, for now. Moreover, assume that one has defined an oriented, orthonormal 3-frame field $\mathbf{e}_i(u^a)$ on the image of *x*, which we denote by $x(\mathcal{O})$.

We shall also assume that if $p = 2$ then the frame field $\mathbf{e}_i(u^a)$ is adapted to the embedding, so one can re-organize the coordinates $x^i$ in such a way that $x(\mathcal{O})$ has coordinates of the form $(x^1, x^2, 0)$ and $\mathbf{e}_a(u) = \partial_b\, h^b_a$, $a, b = 1, 2$, $\mathbf{e}_3 = \partial_3$.

In the case of $p = 3$, *x* might take the form of the inclusion of a subset *U* of $E^3$ in $E^3$; that is, the map $i: U \to E^3$, $x \mapsto x$. Thus, one can factor the map $\mathbf{e} : \mathcal{O} \to SO(E^3)$, $u^j \mapsto (x^i(u^j), \mathbf{e}_i(u^j))$ into the composition of *x*, followed by a map $\mathbf{e} : x(\mathcal{O}) \to SO(E^3)$, $x \mapsto \mathbf{e}_i(x^j)$, which is well-defined, since *x* is assumed to be an embedding.

What we shall be calling a *basic congruence* of curves for the embedding *x* is the set of curves that one gets by fixing all of the parameters $u^a$, except for one, which produces a curve for each value of the remaining parameters; i.e., they are embeddings of the coordinate lines in $\mathbb{R}^p$. Hence, for each $a = 1, \ldots, p$, one can think of the frame field $\mathbf{e}_i(u^a)$ as being an oriented, orthonormal frame field along each curve of that congruence of curves.

If one now looks at the equation for the frame field $\mathbf{e}_i(x^a)$ then one will see that it takes the form:

$$\frac{\partial \mathbf{e}_i}{\partial u^a} = \mathbf{e}_j(u^b) \omega_{ia}^j(u^b). \tag{4.1}$$

Of course, in effect, this means that for each parameter $u^a$, if one holds the other parameters fixed then one will have a system of ordinary differential equations for the frame field along the curves of the basic congruence that is defined by $u^a$. The system will also have the form as (3.15), except that the parameter $s$ becomes $u^a$ and the ordinary derivative becomes a set of partial derivatives.

Similarly, since there are $p$ parameters now, instead of one, there will be a separate infinitesimal rotation matrix:

$$\omega_{ia}^j(u^b) = \begin{bmatrix} 0 & \kappa_a(u) & -\lambda_a(u) \\ -\kappa_a(u) & 0 & \tau_a(u) \\ \lambda_a(u) & -\tau_a(u) & 0 \end{bmatrix} = \tau_a I_1 + \lambda_a I_2 + \kappa_a I_3, \tag{4.2}$$

for each value of $a$. However, one can still interpret the components of that matrix as representing the torsion $\tau_a$ and curvatures $\lambda_a$, $\kappa_a$ of the frame field $\mathbf{e}_i(x^b)$ along the basic congruence that pertains to $a$ as one would for a moving frame along any other curve.

In the case of three-dimensional objects, under the factorization of $\mathbf{e}$ into $\mathbf{e}_i \cdot x$, one sees that, from the chain rule for differentiation, the partial derivatives in (4.1) factor into:

$$\frac{\partial}{\partial u^i} = \frac{\partial x^j}{\partial u^i} \frac{\partial}{\partial x^j}, \tag{4.3}$$

so one can regard the left-hand side of (4.1) as the pull-back of $d\mathbf{e}_i$ by the embedding $x$. Similarly, the right-hand side is also the pull-back of $\mathbf{e}_j \otimes \omega_i^j$ by $x$, and the entire equation is the pull-back of the equation $d\mathbf{e}_i = \mathbf{e}_j \otimes \omega_i^j$, which defines the teleparallelism connection 1-form $\omega_i^j$ when $p = 3$. Thus, we see that for three-dimensional objects the equations of the frame field on $x(\mathcal{O})$ are simply the equations of the teleparallelism connection that it defines, pulled back along the embedding of the object.

The statics of deformation then extend in a straightforward way that mostly follows from adding more parameters to the embedding $x$ and replacing a single ordinary derivative with $p$ partial derivatives.

For instance, when a frame field $\mathbf{e}_i(u)$ on an initial object $x : \mathcal{O} \to SO(E^3)$ is deformed into a frame field $\bar{\mathbf{e}}_i(u)$ on a deformed object $\bar{x} : \mathcal{O} \to SO(E^3)$, the components of the finite frame strain 1-form:

$$\eta_i^j = \eta_{ia}^j \, du^a \tag{4.4}$$

become:

$$\eta_{ia}^j = R_k^j \, \tilde{R}_{i,a}^k = \bar{\omega}_{ia}^j - R_k^j \, \omega_{la}^k \, \tilde{R}_i^l, \tag{4.5}$$

and $\eta_{ia}^j$ takes the form of a set of $p$ skew-symmetric matrices that look like:

$$\eta_{ia}^j = \begin{bmatrix} 0 & \Delta\kappa_a & -\Delta\lambda_a \\ -\Delta\kappa_a & 0 & \Delta\tau_a \\ \Delta\lambda_a & -\Delta\tau_a & 0 \end{bmatrix}. \tag{4.6}$$

Similarly, a finite variation of the initial frame field $\mathbf{e}_i(u)$ will now simply take the form:

$$\Delta\mathbf{e}_i(u^1, \ldots, u^p, h) = \mathbf{e}_i(u^1, \ldots, u^p) \, \tilde{R}(u^1, \ldots, u^p, h), \tag{4.7}$$

which differs from (3.26) only by the number of parameters.

One still differentiates by $h$ at $h = 0$ to get the variation of $\mathbf{e}_i(u)$:

$$\delta\mathbf{e}_i(u) = \frac{\partial(\Delta\mathbf{e}_i)}{\partial h}\bigg|_{h=0} = \mathbf{e}_i(u) \, \delta\tilde{R}(u), \tag{4.8}$$

where:

$$\delta\tilde{R}(u) = \frac{\partial R(u,h)}{\partial h}\bigg|_{h=0} \tag{4.9}$$

now takes the form of a smooth function on $\mathcal{O}$ that takes its values in $\mathfrak{so}(3)$.

The variation of the 1-form $\omega$ that is induced by $\delta R$ is derived in a manner that is analogous to (3.29), except that one increases the number of parameter derivatives. If we suppress the matrix indices $i, j$ in $\omega_{ia}^j$ and $\tilde{R}_i^j$ then we will now have:

$$\delta\omega_a = \frac{\partial(\delta\tilde{R})}{\partial u^a} + [\omega_a, \delta\tilde{R}] \equiv \nabla_\alpha \, \delta\tilde{R}, \tag{4.10}$$

which we see is the pull-back of the corresponding variation:

$$\delta\omega = d(\delta\tilde{R}) + [\omega, \delta\tilde{R}] \equiv \nabla(\delta\tilde{R}) \tag{4.11}$$

of the teleparallelism connection form $\omega$ by the embedding $x$ when the object is three-dimensional.

The couple-stress that is associated with a virtual displacement $\delta \mathbf{e}_i$ of the frame field is now a vector field $\mu_j^i = \mu_j^{ia} \partial_a$ on $\mathcal{O}$ with coefficients in $\mathfrak{so}(3)^*$, and the infinitesimal increment of virtual work that is associated with the virtual displacement $\delta \mathbf{e}_i$ then takes the form:

$$\delta W = \mu_j^{ia} \, \delta \omega_{ia}^j. \tag{4.12}$$

If the couple-stress admits a stress potential $W(\omega_{ia}^j, \omega_{a,b}^j, \ldots)$ then one will have that:

$$\mu_j^{ia} = \frac{\delta W}{\delta \omega_{ia}^j}, \tag{4.13}$$

which will again define a constitutive law for the coupling of infinitesimal frame strain to couple-stress, and in the linear case, one will have an equation of the form:

$$\mu_j^i = C_{jl}^{ik} \omega_k^l, \tag{4.14}$$

which then affects the values that are taken by the vector field $\mu_j^i$ and the 1-form $\omega_i^j$.

The equations of equilibrium have an analogous form to the ones above for the deformable line, namely, (3.35) and (3.36). Once again, the primary difference is in the number of parameters for the embedded object and the associated increased number of derivatives. For a flexible, inextensible line, one will have:

$$0 = \frac{\partial \mu_j^{ia}}{\partial u^a}, \tag{4.15}$$

and when one includes force-stresses, one gets one form of the Cosserat equations:

$$0 = \frac{\partial \sigma_j^a}{\partial u^a}, \qquad 0 = \frac{\partial \mu_{ij}^a}{\partial u^a} + \sigma_i^a x_{ja} - \sigma_j^a x_{ia}. \tag{4.16}$$

Typically, one sees these equations expressed for three-dimensional objects, for which the matrix $x_{,j}^i$ becomes invertible, and if one multiplies by that inverse, the partial derivatives become derivatives with respect to the $x^i$ coordinates, which puts the equations into the form:

$$0 = \frac{\partial \sigma_j^i}{\partial x^i}, \qquad 0 = \frac{\partial \mu_{ij}^k}{\partial x^k} + \sigma_{ij} - \sigma_{ji}, \tag{4.17}$$

which is how they appear most often in the literature.

**5. Discussion.** It is often initially assumed that because teleparallelism involves a connection that must have vanishing curvature, it must somehow be less useful in physical models than the Riemann-Cartan approach to geometry. However, one finds that not only is it common knowledge by now that Einstein's equations of gravitation can be expressed in terms of the torsion tensor for a teleparallelism connection, but as we have see, in continuum mechanics, which is another important class of applications of differential geometry to physics, the way that one imagines curvature and torsion in the context of the deformation of material objects is more related to the equations of moving frames than it is to the obstructions to parallel translation.

In particular, it seems to be common for researchers who consider the inclusion of torsion into the equations of gravitation, such as the Einstein-Cartan-Sciama-Kibble (ECSK) equations, to not see a fundamental mismatch in the coupling of torsion, which relates to infinitesimal translations, to intrinsic angular momentum, which relates to infinitesimal rotations. Of course, this also casts some doubt upon the Einstein equations themselves, which couple curvature, which relates to infinitesimal rotations, to energy-moment-stress, which relates to infinitesimal translations. In that light, one might expect that the teleparallelism form of the equations would be more reasonable, which then suggests that the correct coupling of curvature to spin might have a different sort of character than it takes on in the ECSK picture.

Since both general relativity and continuum mechanics were hindered by the fact that for some time the reigning approach to the geometry of curved spaces was Riemannian geometry, in which torsion plays no role, one might reconsider the possibility that it is not always the most natural geometry to consider in physical models. It was the hope of this preliminary treatment of the application of teleparallelism to continuum mechanics that one might get a better intuition for the place of that kind of geometry in conventional physics.

## References ([1])

---

([1]) References marked with an asterisk are available in English translation at the author's website: www.neo-classical-physics.info .